\documentclass{amsart}

\usepackage{amssymb,amsmath}
\usepackage{verbatim}

\usepackage{tikz}
\usepackage{pgfplots}

\usepackage[section]{algorithm}
\usepackage{algpseudocode}

\usepackage{cleveref}
\usepackage{url}
\usepackage{bm}

\newtheorem{theorem}{Theorem}[section]

\newtheorem{definition}[theorem]{Definition}

\newtheorem{remark}[theorem]{Remark}

\newtheorem{example}[theorem]{Example}

\def\FF{{\mathbb F}}

\def\ie{\hbox{i.e.}}

\def\Gr{Gr\"obner}

\def\GF{{\mathrm{GF}}}

\def\Aradi{{\sc Aradi}}
\def\Trivium{{\sc Trivium}}
\def\Magma{{\sc Magma}}

\def\Reduce{{\textsc{Reduce}}}
\def\GB{{\textsc{Groebner}}}
\def\GBSafe{{\textsc{GroebnerSafe}}}
\def\GBElimLin{{\textsc{GBElimLin}}}

\def\MultiSolve{{\textsc{MultiSolve}}}

\def\maxdeg{{\mathrm{maxdeg}}}
\def\Oracle{{\textsc{Oracle}}}
\def\OracleNRV{{\textsc{OracleNRV}}}
\def\OracleH{{\textsc{OracleH}}}
\def\OracleT{{\textsc{OracleT}}}

\def\tame{{\tt tame}}
\def\wild{{\tt wild}}
\def\status{{\tt status}}

\def\cF{{\mathcal{F}}}

\begin{document}

\title[Oracle-Based Multistep Strategy for Solving $\ldots$] 
{Oracle-Based Multistep Strategy for Solving Polynomial Systems
Over Finite Fields and Algebraic Cryptanalysis of the Aradi Cipher}

\author[R. La Scala]{Roberto La Scala$^*$}

\author[S.K. Tiwari]{Sharwan K. Tiwari$^{**}$}

\address{$^*$ Dipartimento di Matematica, Universit\`a degli Studi di Bari
``Aldo Moro'', Via Orabona 4, 70125 Bari, Italy}
\email{roberto.lascala@uniba.it}

\address{$^{**}$ Cryptography Research Centre, Technology Innovation Institute,
Abu Dhabi, United Arab Emirates}
\email{sharwan.tiwari@tii.ae}

\thanks{
The first author acknowledges the partial support of PNRR MUR projects ``Security and Rights
in the CyberSpace'', Grant ref.~CUP H93C22000620001, Code PE00000014, Spoke 3, and ``National
Center for HPC, Big Data and Quantum Computing'', Grant ref.~CUP H93C22000450007, Code CN00000013,
Spoke 10. The same author was co-funded by MUR project PRIN 2022SC, Grant ref. 2022RFAZCJ, and
Universit\`a di Bari ``Fondo acquisto e manutenzione attrezzature per la ricerca'',
Grant ref.~DR 3191.
}



\begin{abstract}
The multistep solving strategy consists in a divide-and-conquer approach: when a multivariate polynomial
system is computationally infeasible to solve directly, one variable is assigned over the elements of
the base finite field, and the procedure is recursively applied to the resulting simplified systems.
In a previous work by the same authors (among others), this approach proved effective in the algebraic
cryptanalysis of the \Trivium\ cipher.

In this paper, we present a new formulation of the corresponding algorithm based on a Depth-First Search
strategy, along with a novel complexity analysis leveraging tree structures.
We also introduce the notion of an ``oracle function'', which is intended to determine whether evaluating
a new variable is required to simplify the current polynomial system. This notion allows us
to unify all previously proposed variants of the multistep strategy, including the classical hybrid approach,
by appropriately selecting the oracle function.

Finally, we employ the multistep solving strategy in the cryptanalysis of the NSA's recently introduced low-latency
block cipher \Aradi, achieving a first full-round algebraic attack that exposes structural features in its symbolic
model.
\end{abstract}

\maketitle


\section{Introduction}

Among NP-hard problems, one with immediate and wide-ranging applications is the so-called MP-problem,
which involves solving a system of multivariate polynomial equations over a finite field
$\FF = \GF(q)$. When $q = 2$, this problem corresponds to the SAT-problem, and many challenges
in cryptography are directly related to the MP-problem, including algebraic cryptanalysis
and multivariate public-key cryptography.

Among the methods used to solve the MP-problem are symbolic algorithms, such as \Gr\ bases and XL
linearization \cite{CKPS,F4,F5}, SAT solvers \cite{MM} for the Boolean case and discrete optimization
techniques like Quantum Annealing \cite{QA}. Each of these methods can be advantageous
in specific cases, but in general, they all exhibit a complexity comparable to the exponential
complexity $q^n$ of the brute-force approach, with $n$ representing the number of variables
in the polynomial system.

Beyond theoretical worst-case complexity, valid when the polynomials in the system are randomly
chosen, it is especially important in cryptanalysis to estimate the actual computational cost
of solving a specific polynomial system over a finite field through experimentation.
Since solving such systems becomes infeasible when the number of variables is sufficiently
large, a standard approach is to combine brute force with polynomial solving in what is known
as a ``hybrid strategy''. Specifically, this approach assumes that one can feasibly solve
all derived systems obtained by exhaustively assigning values in the finite field to $k$
out of the $n$ variables in the original system. By using a test set of such partial evaluations,
it is possible to estimate the average solving time $\tau$ for a single derived system.
Consequently, the total expected complexity of the hybrid approach becomes $q^k\tau$.

One issue with this approach is that, in order to ensure the actual solvability of all
derived systems, the number $k$ of evaluated variables may become quite large. To address this,
the paper \cite{LSPTV2} proposes a dynamic approach in which the number of evaluated variables
is adjusted based on the specific evaluation. In other words, starting from an initial evaluation
$x_1 = c_1,\ldots,x_k = c_k\ (c_1,\ldots,c_k\in\FF$), a set of $q$ additional assignments
$x_{k+1} = c_{k+1}$, for all $c_{k+1}\in\FF$, is considered, but only when the polynomial
system associated with the initial evaluation exhibits features that indicate it may be
difficult to solve. This divide-and-conquer strategy is referred to as a ``multistep strategy''.

In \cite{LSPTV2}, the key feature used to predict the actual solvability of the system was the number
of variables remaining after a preprocessing step, which was required to be below a given threshold. 
This step aimed to generate additional linear polynomials in the ideal, beyond those corresponding
to variable assignments, in order to eliminate as many variables as possible. The preprocessing
was based on the computation of a truncated \Gr\ basis up to a low degree.

Since the number of remaining variables does not yield accurate predictions in the case of sparse
polynomial systems encountered in the algebraic cryptanalysis of certain block ciphers, in this
paper we introduce an abstract notion of an ``oracle function''. This function determines whether
the multistep strategy should attempt to solve the current system or instead perform a branching
operation by assigning a new variable across the elements of the finite field. The recursive calls
in a Depth-First Search implementation of the multistep strategy naturally define a tree structure,
whose analysis provides insights into the algorithm’s complexity. The concept of oracle
function provides a unified framework that generalizes different solving strategies, including
the one used in \cite{LSPTV2} and the classic hybrid approach, into a single algorithm in which only
the oracle function varies. Furthermore, it is plausible that machine learning techniques may aid
in the construction of effective oracle functions tailored to specific problem settings.
Finally, this paper demonstrates the applicability of the multistep solving strategy through
a full-round algebraic attack on a modern block cipher.

Low-latency block ciphers have gained prominence in applications such as secure memory encryption,
where high throughput and minimal critical path delay are essential. \Aradi, a 128-bit block
cipher with a 256-bit key, was recently proposed in \cite{GMW} as a candidate to meet these demands.
\Aradi\ employs a substitution-permutation network (SPN) architecture composed of word-wise
Toffoli gates, word permutations, and lightweight linear maps.
Despite its promising efficiency, the original specification offers limited justification
for the cipher's structural design and includes only a minimal security analysis.

Following the public release of \Aradi\ and its authenticated encryption mode LLAMA by NSA
researchers Greene, Motley, and Weeks \cite{GMW}, several studies have explored
its cryptographic strength and structural properties. One of the earliest responses came
from Avanzi, Dunkelman, and Ghosh \cite{ADG}, who questioned \Aradi's practical advantage
over existing lightweight ciphers in terms of area-latency trade-offs. They also identified
issues in the LLAMA mode related to the use of variable-length initialization vectors,
which compromised the integrity and confidentiality of ciphertexts. These issues were later
acknowledged by the designers in an updated ePrint version.

The first comprehensive cryptanalytic evaluation was conducted by Bellini et al.~\cite{CLAASPING-ARADI}
using the CLAASP framework for automated trail-based analysis. Their study includes statistical
black-box testing (such as avalanche and diffusion analysis), the discovery of differential and
linear trails up to 9 and 10 rounds, impossible differentials up to 8 rounds, and neural
distinguishers up to 5 rounds. A preliminary algebraic analysis is also presented, revealing
an integral distinguisher for 7 rounds requiring $2^{124}$ data, along with an 11-round key-recovery
attack based on a 9-round differential trail, with data and time complexities of $2^{126}$
and $2^{252}$, respectively. These results indicate slow diffusion and weak mixing in the early
rounds of \Aradi.

Another study by Bellini et al.~\cite{BRRT} focuses on the algebraic structure
of the Toffoli-based S-box layer in \Aradi. The authors identify a structural property that
allows any $r$-round integral distinguisher to be extended to $(r+1)$-rounds without increasing
data complexity, based on the predictable interaction of bitwise Toffoli gates. They show that
this phenomenon is preserved under various changes to the linear layer and Toffoli gate ordering.
The paper presents 8-round algebraic distinguishers and a 10-round key-recovery attack requiring
$2^{124}$ data and $2^{177}$ time.

The structural property was first analyzed in a preliminary ePrint report \cite{BRRT-Eprint},
where MILP-based modeling of the three-subset division property without unknown subsets was applied
to \Aradi\ to derive degree bounds. Building on this foundation, a follow-up study
\cite{KKLHSH} extended the analysis to 8 rounds by refining the MILP constraint
formulation introduced in \cite{BRRT-Eprint}. This extension further confirmed
the persistence of the structural weakness across various constraint encoding strategies.

In this work, we study the cryptanalytic properties of \Aradi\ using a symbolic algebraic approach.
Specifically, we derive an implicit representation of the cipher’s round function as a multivariate
polynomial map over the Boolean field $\FF = \GF(2)$. This includes a full symbolic description
of key addition, the nonlinear S-box layer, and the word-wise linear transformation, as defined
in the cipher’s specifications \cite{GMW}. Using this symbolic model, we instantiate the
{\em full 16-round} cipher with a known plaintext–ciphertext pair to obtain a multivariate
polynomial system over $\FF$, whose unknowns correspond to the 256 key bits and the internal
variables introduced by the S-box layers.

To solve this polynomial system over the base field $\FF$, we employ a multistep strategy that explores
the cipher’s algebraic structure. Specifically, it recursively considers partial key assignments
using a Depth-First Search approach. The approach integrates a timeout-based \Gr\ basis
subroutine, used to probe whether the resulting polynomial system is feasibly solvable.

We conduct comprehensive experiments to evaluate the efficiency and reliability of the proposed
method. Specifically, we compare the multistep strategy with the classical hybrid approach
by measuring the success rate of computing a complete \Gr\ basis under a fixed solving degree
and within a predefined time limit, across a test set involving the evaluation of different
numbers of variables. Our results show that in approximately 99.98\% of cases, the polynomial
systems can be efficiently solved after guessing just 251 out of 256 key bits. For the remaining
0.02\%, the multistep strategy automatically switches to 252 evaluations.
Consequently, this strategy achieves a somewhat higher efficiency than
the classic hybrid method, which requires 252 guesses to ensure that 100\%
of systems can be solved efficiently.

Moreover, we observe that all derived systems are quadratic and can be solved with a solving degree
of two when more than 250 key variable assignments are provided. However, this is no longer possible
when the number of assignments is reduced to 250. These findings point to some thresholds within
\Aradi’s algebraic structure, and they show that a dynamic multistep strategy can be practically
useful in algebraic cryptanalysis.


\section{Multistep algorithm}

Let $\FF = \GF(q)$ be a finite field and consider the multivariate polynomial algebra $R = \FF[x_1,\ldots,x_n]$
and the field equation ideal $\cF = \langle x_1^q - x_1,\ldots,x_n^q - x_n \rangle \subset R$.
Let $J = \langle f_1,\ldots,f_m\rangle \subset R$ be an ideal such that $\# V_\FF(J)\leq 1$.
Put
\[
H = \{f_1,\ldots,f_m\} \cup \{x_1^q - x_1,\ldots,x_n^q - x_n\}
\]
that is, $H$ is a generating set of the ideal $J + \cF$. To simplify notations, we denote
$\FF = \{c_1,\ldots,c_q\}$. Since $\# V_\FF(J)\leq 1$, we have that a (reduced) \Gr\ basis
of $J + \cF$ with respect to any monomial ordering of $R$, is either $G = \{1\}$ or
$G = \{x_1 - c_{i_1},\ldots,x_n - c_{i_n}\}$, for some $1\leq i_1,\ldots,i_n\leq q$
(see, for instance, \cite{GLS,LSPTV,LST}). In other words, $\maxdeg(G) = \max\{\deg(f)\mid f\in G\}\leq 1$.

Let $d_1\geq 0$ be an integer and denote by $\GB(H, d_1)$ the algorithm performing an incomplete
\Gr\ basis computation which is stopped when all S-polynomials of degree $\leq d_1$
have been consider at a current step. This algorithm corresponds having a bound on the size
of the Macaulay matrix when using linear algebra methods for computing the \Gr\ basis.
If the degree $d_1$ is sufficiently large, we have that $G = \GB(H, d_1)$ is a complete
\Gr\ basis of $J + \cF$. In this case, we say that $d_1$ is a {\em solving degree} for the ideal
$J + \cF$ with respect to the given monomial ordering. However, if $d_1 < \maxdeg(H)$,
then $G = \GB(H, d_1)$ is generally not a generating set of the ideal $J + \cF$. Instead,
a generating set is given by $G \cup G'$, where $G' = \{g\in H\mid \deg(g) > d_1\}$.

A useful procedure within the solving algorithm is \GBElimLin. Its goal is
to generate linear polynomials belonging to the ideal $J + \cF$ in order to eliminate variables
from its generators. The method consists of computing an incomplete \Gr\ basis up to a sufficiently
low degree $d_1$ to keep the computation efficient. The expectation that linear polynomials appear
among the generators arises from the fact that the complete \Gr\ basis is entirely linear when
$\# V_\FF(J) = 1$. Note also that, when incorporating this procedure into the multistep solving
algorithm, we explicitly add linear polynomials corresponding to variable evaluations to the generating set.

\suppressfloats[b]
\floatname{algorithm}{Algorithm}
\begin{algorithm}
\caption{\GBElimLin}
\begin{tabular}{l@{\ }l}
\hspace{-10pt} {\bf Input}: & a \hfill generating \hfill set \hfill $H$ \hfill of \hfill the \hfill ideal
\hfill $J + \cF$ \hfill and \hfill an \hfill integer \hfill $d_1\geq 0$; \\
\hspace{-10pt} {\bf Output}: & two \hfill subsets \hfill $L,G\subset J + \cF$ \hfill where \hfill $\deg(g) = 1$
\hfill for \hfill all \hfill $g\in L$ \hfill and \\
& $\langle G\rangle = (J + \cF)\cap R'$ with $R'$ the polynomial algebra in the variables \\
& occuring in $G$.
\end{tabular}
\begin{algorithmic}[0]
\State $G:= \{g\in H\mid \deg(g)\leq d_1\}$;
\State $G':= H\setminus G$;
\State $G:= \GB(G, d_1)$;
\State $G:= G\cup G'$;
\If{$\maxdeg(G)\leq 1$}
\State \Return $\emptyset, G$;
\EndIf;
\State $L:= \{g\in G\mid \deg(g) = 1\}$;
\State $G':= G\setminus L$;
\State $S:= \{x_i^q - x_i\mid x_i\ \mbox{occurs in}\ G\}$;
\State $G:= \Reduce(G', L \cup S)$;
\State $S:= \{x_i^q - x_i\mid x_i\ \mbox{occurs in}\ G\}$;
\State $G:= G\cup S$;
\State \Return $L,G$;
\end{algorithmic}
\end{algorithm}

\newpage
In the \GBElimLin\ algorithm, we assume that the linear polynomials in the set $L$ are fully
interreduced, meaning they are in reduced row echelon form in the Macaulay matrix used
to compute $\GB(G, d_1)$.
The \Reduce\ procedure implements the full reduction algorithm from \Gr\ basis theory.
Note that if $\maxdeg(G)\leq 1$, then $G$ is a \Gr\ basis of $J + \cF$. In other words,
\GBElimLin\ directly returns $V_\FF(J)$, as $d_1$ is a solving degree in this case.
Finally, note that a modified version of \GBElimLin\ could be iteratively applied to enhance
its capability to eliminate variables.

To describe our solving algorithm, it is also helpful to consider a variant of \GB\ with
a timeout $\tau$ to limit the computation time. Specifically, we define the following procedure.

\suppressfloats[b]
\floatname{algorithm}{Algorithm}
\begin{algorithm}
\caption{\GBSafe}
\begin{tabular}{l@{\ }l}
\hspace{2pt} {\bf Input}: & a \hfill generating \hfill set \hfill $H$ \hfill of \hfill an \hfill ideal \hfill $J + \cF$,
an \hfill integer \hfill $d_2\geq\maxdeg(H)$ \hfill and \\
& a real number $\tau > 0$; \\
\hspace{2pt} {\bf Output}: & either \hfill $\wild, \emptyset$ \hfill or \hfill $\tame, G$ \hfill where \hfill
$G = \GB(H, d_2)$ \hfill and \\
& $\maxdeg(G)\leq 1$; \\
\end{tabular}
\begin{algorithmic}[0]
\State {\bf compute} $G:= \GB(H, d_2)$ within timeout $\tau$;
\If{computation completes within $\tau$ and $\maxdeg(G)\leq 1$}
\State \Return $\tame, G$;
\EndIf
\State \Return $\wild,\emptyset$;
\end{algorithmic}
\end{algorithm}

\newpage
We finally present the solving algorithm \MultiSolve. To describe it, we need to introduce
the following subsets of $R$. For all $1\leq i_1,\ldots,i_k\leq q$ and $0\leq k\leq n$, we put
\[
E(i_1,\ldots,i_k) = \{ x_1 - c_{i_1}, \ldots, x_k - c_{i_k} \}.
\]
For $k = 0$, we put $E() = \emptyset$ by definition. Adding the subset $E(i_1,\ldots,i_k)$
to the generators of the ideal $J$ corresponds to evaluating the variables $x_1,\ldots,x_k$
at the elements $c_{i_1},\ldots,c_{i_k}\in\FF$. More precisely, we define the ideal
\[
J(i_1,\ldots,i_k) = \langle E(i_1,\ldots,i_k) \rangle + J.
\]
If $V_\FF(J) = \{(c_{i_1},\ldots,c_{i_n})\}$, we have that
\[
V_\FF(J(j_1,\ldots,j_k)) =
\left\{
\begin{array}{cl}
V_\FF(J) & \mbox{if}\ (j_1,\ldots,j_k) = (i_1,\ldots,i_k); \\
\emptyset & \mbox{otherwise}.
\end{array}
\right.
\]

The \MultiSolve\ algorithm essentially follows a divide-and-conquer approach. If computing
$V_\FF(J(i_1,\ldots,i_k))$ within a given time limit is not feasible, the algorithm instead
attempts to compute $V_\FF(J(i_1,\ldots,i_k,i))$, for all $1\leq i\leq q$. Indeed, we have
\[
V_\FF(J(i_1,\ldots,i_k)) = \bigcup_{1\leq i\leq q} V_\FF(J(i_1,\ldots,i_k,i))
\]
Moreover, the generators of $J(i_1,\ldots,i_k,i)$ involve fewer unknowns than those
of $J(i_1,\ldots,i_k)$. In other words, this approach simplifies the solving problem
at the cost of branching over the elements of the base finite field. 

The algorithm \MultiSolve\ is implemented here as a recursive procedure of type Depth-First
Search (DFS), which is alternative to the iterative algorithm of type Breadth-First Search (BFS)
that was proposed in the paper \cite{LSPTV2}.

Let $H$ be a generating set of the ideal $J + \cF$ such that $\# V_\FF(J)\leq 1$. Let \Oracle\ be
a function that takes as input a set of polynomials and an integer $k$, and returns either the string
\tame\ or the string \wild. Finally, fix two integers $d_1\geq 0$ and $d_2\geq\maxdeg(H)$. The set $H$,
along with the function $\Oracle$ and the integers $d_1,d_2$, are global variables for the following
algorithm \MultiSolve\ which applies the multistep solving strategy to the ideal $J(i_1,\ldots,i_k)$.
For $k = 0$, we have $J() = J$ by definition.

\suppressfloats[b]
\floatname{algorithm}{Algorithm}
\begin{algorithm}
\caption{\MultiSolve}
\begin{tabular}{l@{\ }l}
\hspace{-36pt} {\bf Input}: & a $k$-tuple $(i_1,\ldots,i_k)$ where $1\leq i_1,\ldots,i_k\leq q$
and $0\leq k\leq n$; \\
\hspace{-36pt} {\bf Output}: & a \Gr\ basis $G$ of the ideal $J(i_1,\ldots,i_k) + \cF$; \\
\end{tabular}
\begin{algorithmic}[0]
\State $G:= E(i_1,\ldots,i_k)\cup H$;
\State $L, G:= \GBElimLin(G, d_1)$;
\If{$\maxdeg(G) = 1$}
\State \Return $L\cup G$;
\EndIf
\If{$\maxdeg(G) = 0$}
\State \Return $\{1\}$;
\EndIf
\If{$\Oracle(G, k) = \tame$}
\State $\status, G:= \GBSafe(G, d_2, \tau)$;
\If{$\status = \tame$ and $\maxdeg(G) = 1$}
\State \Return $L\cup G$;
\EndIf
\If{$\status = \tame$ and $\maxdeg(G) = 0$}
\State \Return $\{1\}$;
\EndIf
\EndIf
\ForAll{$i\in\{1,2,\ldots,q\}$}
\State $G':= \MultiSolve(i_1,\ldots,i_k,i)$;
\If{$\maxdeg(G') = 1$}
\State \Return $G'$;
\EndIf
\EndFor
\State \Return $\{1\}$;
\end{algorithmic}
\end{algorithm}

\newpage
Besides the DFS implementation, a key difference of the above algorithm from the one
presented in \cite{LSPTV2} is the use of a general oracle function meant to predict when
$\GBSafe(G, d_2, \tau)$ would return $\status = \wild$. In the previous version of \MultiSolve,
given an integer $B > 0$, the oracle function was simply defined as
\[
\OracleNRV(G, k) =
\left\{
\begin{array}{ll}
\wild & \mbox{if the number of remaining variables in $G$ is $> B$}; \\
\tame & \mbox{otherwise}.
\end{array}
\right.
\]
By remaining variables above, we refer to the variables that still occur in $G$ after applying
the \GBElimLin\ procedure. Indeed, an accurate oracle predicting wild cases could eliminate
unnecessary and time-consuming computations, significantly enhancing the efficiency of the algorithm.
Although the number of variables plays a crucial role in determining the complexity of solving
a polynomial system, many other features can also influence it, especially in the sparse case.

The termination of the \MultiSolve\ algorithm follows directly from the fact that computing
a \Gr\ basis for the ideal $J(i_1,\ldots,i_n) + \cF$ is trivial. This is because all variables
in $R$ are explicitly evaluated through the set
\[
E(i_1,\ldots,i_n) = \{ x_1 - c_{i_1}, \ldots, x_n - c_{i_n} \}.
\]
Of course, termination typically occurs well before evaluating all variables.

We remark that the \MultiSolve\ algorithm essentially imposes a fixed solving degree $d_2$,
at the cost of performing branching among the various derived polynomial systems. Combined with
its recursive structure, this provides \MultiSolve\ with considerable simplicity and flexibility,
which can be adjusted by varying the parameters $d_1,d_2,\tau$ and the function \Oracle.


\section{Complexity}

Let us analyze the complexity of \MultiSolve\ when computing $V_\FF(J(i_1,\ldots,i_k))$.
Recall that for $k = 0$, we have $J() = J$. Indeed, we assume that the computation time
of \GBElimLin\ is negligible, due to the low value of the parameter $d_1$.
Likewise, we assume that invoking the \Oracle\ function is negligible, as these functions
merely involve comparing numerical data or performing a forward pass of a neural network,
operations that are typically very fast.

As a result, the execution time of \MultiSolve\ is the sum of the time spent by \GBSafe\
handling wild and tame cases when $\Oracle = \tame$. We refer to cases corresponding to
the status returned by \GBSafe\ as {\em computed wild} and {\em computed tame} cases.
Similarly, cases determined by the values of the function \Oracle\ are referred to
as {\em predicted wild} and {\em predicted tame} cases.

By a {\em case} or a {\em guess}, we mean here a $l$-tuple $(j_1,\ldots,j_l)$
($1\leq j_1,\ldots,j_l\leq q$, $k\leq l\leq n$) which defines the evaluation set
\[
E(j_1,\ldots,j_l) = \{ x_1 - c_{j_1}, \ldots, x_l - c_{j_l} \}.
\]
These guesses arise during the execution of \MultiSolve\ and correspond to its recursive calls.
The following two notions are useful for the complexity analysis of \MultiSolve.

\begin{definition}
The function \Oracle\ is called {\em accurate} if no computed wild cases occur during the execution
of \MultiSolve. It is called {\em perfect} if, in addition, it minimizes the number of computed tame
cases. In other words, an oracle is accurate if every predicted tame case corresponds to a computed
tame case, and it is perfect if the prediction and computation always agree.
\end{definition}

By assuming that the oracle function is accurate, the complexity of \MultiSolve\ is reduced to the number
of computed tame cases. In other words, there are no computed wild cases incurring the timeout cost $\tau$,
and consequently no significative overhead arises in dealing with wild and tame cases according to
the multistep solving strategy. Indeed, the oracle function associated with the \MultiSolve\ algorithm
in \cite{LSPTV2} is assumed to be accurate. Moreover, given an integer $B > 0$, the standard hybrid
strategy, as described in \cite{BFP1, BFP2}, can be obtained as the \MultiSolve\ algorithm implementing
an accurate oracle function of type
\[
\OracleH_B(G, k) =
\left\{
\begin{array}{ll}
\wild & \mbox{if $k < B$}; \\
\tame & \mbox{otherwise}.
\end{array}
\right.
\]
Another useful concept is the following.

\begin{definition}
A {\em $q$-ary tree} is by definition a tree in which each node has either zero or exactly $q$ children.
In other words, every internal node has $q$ children, while leaf nodes have no children.
\end{definition}

Note that the recursive calls of the \MultiSolve\ algorithm form a $q$-tree, where each node represents
a guess $(j_1,\ldots,j_l)$. Internal nodes correspond to wild cases, whether predicted or computed, while
the leaves represent the computed tame cases. More precisely, we have a complete $q$-tree when
$V_\FF(J(i_1,\ldots,i_k)) = \emptyset$. Conversely, when $V_\FF(J(i_1,\ldots,i_k))\neq\emptyset$,
the tree consists of only a subset of nodes, as recursive calls are terminated as soon as the solution
is found. We have the following useful result concerning $q$-trees.

\begin{theorem}
\label{tree}
In a $q$-ary tree, let $N$ be the total number of nodes, $M$ the number of internal nodes
and $L$ the number of leaves. Then, the following equations holds
\[
N = \frac{q L - 1}{q - 1}, M = \frac{L - 1}{q - 1}.
\]
In particular, when $q = 2$, se have $N = 2L - 1$ and $M = L - 1$.
\end{theorem}

\begin{proof}
We prove the first formula by induction on the number of leaves $L$. When there is only one leaf,
the tree consists of a single node. In this case, we have $N = L = 1$, which satifies the formula.

Assume that the first formula holds for any $q$-ary tree with strictly fewer than $L$ leaves.
Consider a $q$-ary tree $T$ having $N$ nodes and $L$ leaves. This tree can be obtained by
a smaller tree $T'$ by converting one of its leaves into an internal node and attaching $q$ new
leaves to it. Let $N'$ and $L'$ be the number of nodes and leaves of $T'$, respectively.
Then, we have
\[
N = N' + q, L = (L' - 1) + q.
\]
Since $L > L'$, we can apply the induction hypothesis to $T'$, yielding
\[
N = \frac{q L' - 1}{q - 1} + q = \frac{q (L + 1 - q) - 1}{q - 1} + q =
\frac{q L - 1}{q - 1}.
\]
The second formula is obtained directly from the first by subtracting $L$.
\end{proof}

\begin{definition}
During the execution of the algorithm $\MultiSolve(i_1,\ldots,i_k)$, we call {\em depth}
the current number of evaluated variables in addition to $x_1,\ldots,x_k$. Thus, the algorithm
starts at depth 0 and reaches a maximum depth $l$, meaning that the largest set of evaluated
variables is $\{x_1,\ldots,x_k,x_{k+1},\ldots,x_{k+l}\}$. In other words, the depth corresponds
to the current number of recursion levels.
\end{definition}

Let $\Oracle$ be any function and assume that the depth of computed tame cases ranges
from a minimum of $B$ to a maximum of $C$. The \MultiSolve\ algorithm implementing $\OracleH_C$,
which corresponds to the standard hybrid strategy, computes exactly $q^C$ tame cases.
Since it is an accurate oracle, we have no computed wild cases and the number of predicted wild
cases is
\[
1 + q + \ldots + q^{C-1} = \frac{q^C - 1}{q - 1}.
\]

\begin{theorem}
The total numbers of tame and wild cases corresponding to any oracle function, are bounded
above by the corresponding numbers for $\OracleH_C$.
\end{theorem}

\begin{proof}
Consider the $q$-tree associated with the recursive calls of the \MultiSolve\ algorithm when
implementing a general function \Oracle. If $B = C$, meaning that all tame cases occur
at depth $C$, then the total number of tame and wild cases is exactly the same as for
$\OracleH_C$. If $B < C$, there exists a tame case at some depth $B\leq r < C$.
In this case, the $q$-tree corresponding to \Oracle\ excludes the $q^{C-r}$ tame cases
at depth $C$ that would have originated from that node in the $q$-tree of $\OracleH_C$.
Similarly, the corresponding wild cases, whose total number is
\[
1 + q + \ldots + q^{C-r-1} = \frac{q^{C-r} - 1}{q - 1}
\]
are also excluded from the $q$-tree defined by \Oracle.
\end{proof}

When the oracle function is accurate, such as \OracleNRV\ in the attack on \Trivium\ \cite{LSPTV2},
all computational effort is devoted, up to negligible overhead, to solving the tame cases, and the
above result implies the superiority of the multistep strategy over the hybrid one.
However, if \OracleNRV\ is not applicable, for instance in the sparse case where the number
of remaining variables is insufficient to accurately predict the output status of \GBSafe\
computations, different features of the input polynomial systems has to be considered to design
an accurate oracle.

A promising approach we are currently pursuing is the use of machine learning techniques
to build a data-driven oracle for predicting the wall-clock runtime of the Block-Wiedemann procedure
when applied to sparse Macaulay matrices arising in the multistep process. We adopt a permutation-invariant
Deep Sets architecture \cite{ZKRPSS}, which processes each row independently, aggregates the resulting
embeddings to enforce invariance with respect to row ordering, and combines them with coarse-grained metadata.
The resulting global representation is then passed to a final regression head that outputs a scalar
runtime prediction. The model is trained with mean squared error loss and evaluated through repeated
cross-validation, assessing both predictive accuracy and how runtime relates to sparsity, matrix
dimensions, and construction parameters. These results will be reported in a forthcoming paper.

The following is a trivial, non-predictive oracle that is always available
\[
\OracleT(G, k) = \tame.
\]
Indeed, this constant tame function implies that all wild cases are computed ones. By definition,
a perfect oracle function has the same computed tame cases as \OracleT\ and no computed
wild cases. In other words, for a fixed choice of the parameters $d_1,d_2$ and $\tau$,
a perfect oracle achieves optimal \MultiSolve\ performance in terms of the number of \GBSafe\
executions, that is, computed wild and tame cases. At the opposite end of the spectrum,
the function \OracleT\ yields the worst performance.

We now compare the worst-case and optimal oracles, showing that the overhead introduced by computed
wild cases remains moderate, especially when compared with the high complexity levels encountered
in cryptography. Applying Theorem \ref{tree} then yields the following result.

\begin{theorem}
Let $N$ denote the number of \GBSafe\ executions in the \MultiSolve\ algorithm using \OracleT,
and let $L$ be the corresponding number when using a perfect oracle function. The resulting
speed-up ratio $N/L$ is given by
\[
\frac{N}{L} = 1 + \frac{1 - 1/L}{q-1}
\]
and therefore
\[
1 \le \frac{N}{L} < \frac{q}{q-1}.
\]
In particular, for $q = 2$, we have $N/L = 2 - 1/L$ and $1\leq N/L < 2$.
\end{theorem}

\begin{proof}
Recall that the recursive calls of \MultiSolve\ define a $q$-ary tree. For the function \OracleT,
these calls correspond exactly to executions of \GBSafe, and therefore their number $N$ can be
expressed as
\[
N = \frac{q L - 1}{q - 1}
\]
where $L$ is the number of computed tame cases. This $L$ also represents the total number of \GBSafe\
executions when using a perfect oracle. Computing the ratio $N/L$ yields
\[
\frac{N}{L} = \frac{q L - 1}{L(q - 1)} = 1 + \frac{1 - 1/L}{q-1}
\]
To establish the bounds, observe that the function
\[
f(L) = 1 + \frac{1 - 1/L}{q-1}
\]
is strictly increasing for $L\geq 1$, since $f'(L) = \frac{1}{(q-1)L^2} > 0$.
Hence, the minimum of $f$ over $L\geq 1$ is attained at $L = 1$, giving
\[
\min_{L\geq 1} \frac{N}{L} = f(1) = 1.
\]
Moreover, since $1 - 1/L < 1$ for all finite $L > 1$, we obtain the strict upper bound
\[
\frac{N}{L} = f(L) < 1 + \frac{1}{q-1} = \frac{q}{q-1}.
\]
\end{proof}

Since each \GBSafe\ call, whether corresponding to a tame or a wild case, is bounded
by time $\tau$, the total running time using \OracleT\ is at most a factor of $\frac{q}{q-1}$
larger than with a perfect oracle. In particular, for $q = 2$ this overhead is strictly
less than $2$, corresponding to at most a one-bit increase in the exponent of the complexity,
while for large field sizes $q$ it becomes negligible as $N/L \to 1$.

\medskip
Note that choosing a constant wild oracle function reduces \MultiSolve\ to a recursive
application of \GBElimLin. For $d_1 = 0$, this corresponds to brute-force enumeration
of all variable assignments over the base finite field $\FF$.

From a practical perspective, by applying \OracleT\ to a suitable test set of ideals
$J(i_1,\ldots,i_k)$, we can estimate the minimal and maximal depths for the computed tame cases,
denoted by $B$ and $C$ respectively, for other similar ideals. Indeed, since we aim to
compute $V_\FF(J)$ and
\[
V_\FF(J) = \bigcup_{1\leq i_1,\ldots,i_k\leq q} V_\FF(J(i_1,\ldots,i_k))
\]
it is necessary to apply \MultiSolve\ to all ideals $J(i_1,\ldots,i_k)$.
The classical hybrid strategy consists of using the oracle function $\OracleH_C$.
If a perfect or, at least, an accurate oracle function is not available,
a first approximation can be provided by $\OracleH_B$. Note that using this oracle
actually corresponds to executing \MultiSolve\ with the function \OracleT, for all ideals 
$J(i_1,\ldots,i_k,i_{k+1},\ldots,i_{k+B})$ where $1\leq i_{k+1},\ldots,i_{k+B}\leq q$.
A more refined oracle function may be derived from a detailed analysis of the distribution
of wild and tame cases, for example using machine learning techniques.


\section{An Overview of \Aradi}

Henceforth, we assume the base field to be the Boolean field $\FF = \GF(2)$.
Aradi\ is a low-latency block cipher proposed by researchers at the United States National Security Agency (NSA),
introduced by Greene et al.~\cite{GMW}, to support secure memory encryption applications.

It operates on 128-bit blocks and uses a 256-bit key. The cipher follows a substitution-permutation network (SPN)
design. Each round consists of three steps: key addition, a nonlinear substitution layer $\pi$, and a linear
diffusion layer $\Lambda_i$ ($0\leq i\leq 3$). The full encryption process consists of 16 rounds followed
by a final key addition, and is expressed as the following composition of functions:
\begin{equation*}
    \tau_{RK^{16}} \circ \bigcirc_{i = 0}^{15} (\Lambda_{i \bmod 4} \circ \pi \circ \tau_{RK^i}),
\end{equation*}
where $\tau_{RK^i}$ denotes XOR with the $i$-th round key $RK^i$ and composition $\circ$ is read
from right to left.

The internal state is represented by four 32-bit words $(W, X, Y, Z)$.

\subsection{Substitution Layer $\bm{\pi}$}

The substitution layer introduces nonlinearity via bitwise Toffoli gates across the word-level structure.
The transformation $\pi$ operates on three words at a time, following the update rule
$(a, b, c) \mapsto (a, b, c \oplus (a \wedge b))$. Note that we use standard Boolean notation, where $\oplus$
denotes the XOR operation, \ie, addition in $\FF = \GF(2)$, and $\wedge$ denotes the AND operation,
corresponding to multiplication in $\FF$. Applied to the full state, $\pi$ updates the words as:
\begin{align*}
    X &\leftarrow X \oplus (W \wedge Y), \nonumber \\
    Z &\leftarrow Z \oplus (X \wedge Y), \nonumber \\
    Y &\leftarrow Y \oplus (W \wedge Z), \nonumber \\
    W &\leftarrow W \oplus (X \wedge Z).
\end{align*}
This structure can be interpreted as the parallel application of 32 instances of a 4-bit S-box, each defined
by a cascade of Toffoli gates, applied independently to each bit position across the words.

\subsection{Linear Layer $\bm{\Lambda_i}$}

The linear layer enhances diffusion and is determined by the round index modulo 4. A linear transformation
$L_i$ is applied independently to each of the four 32-bit state words:
\[
\Lambda_i(W, X, Y, Z) = (L_i(W), L_i(X), L_i(Y), L_i(Z)).
\]
Each map $L_i$ is an involutive function defined on 32-bit words, where each word is viewed as a pair
of 16-bit halves $(u, l)$. Namely, the transformation $L_i$ is defined as follows:
\begin{align*}
(u, l) \mapsto 
\Big(u \oplus S^{a_i}_{16}(u) \oplus S^{c_i}_{16}(l),\
l \oplus S^{a_i}_{16}(l) \oplus S^{b_i}_{16}(u)\Big),
\end{align*}
where $S^{m}_{16}$ denotes a left circular shift by $m$ bits on a 16-bit word. The constants $(a_i, b_i, c_i)$
are determined based on the round index $i \bmod 4$, as specified in Table~\ref{table:shifts}.

\begin{table}[!htb]
 	\centering 
 	\caption{Shift offsets in the linear map $\Lambda_i$}
 	\label{table:shifts}
 	\begin{tabular}{c | c c c}
 		$i$ & $a_i$ & $b_i$ & $c_i$ \\
		\hline
 		0 & 11 & 8  & 14 \\
 		1 & 10 & 9  & 11 \\
 		2 & 9  & 4  & 14 \\
 		3 & 8  & 9  & 7
 	\end{tabular}
\end{table}

\subsection{Key Addition}
Each round begins with XORing a 128-bit round key to the current state. That is,
\[
(W, X, Y, Z) \leftarrow (W \oplus RK^i_0, X \oplus RK^i_1, Y \oplus RK^i_2, Z \oplus RK^i_3),
\]
where \((RK^i_0, RK^i_1, RK^i_2, RK^i_3)\) are the four 32-bit words that constitute the $i$-th round key,
as defined in the key schedule depending on the parity of $i$.

\subsection{Key Schedule}

At step $i = 0$, the 256-bit master key is divided into eight 32-bit registers, denoted as
$(K_0^0, K_1^0, \ldots, K_7^0)$.

At each step $i \geq 0$, the round key $RK^i$ is defined as follows:
\[
RK^i =
\begin{cases}
K_0^i \Vert K_1^i \Vert K_2^i \Vert K_3^i & \text{if } i \text{ is even}, \\
K_4^i \Vert K_5^i \Vert K_6^i \Vert K_7^i & \text{if } i \text{ is odd}.
\end{cases}
\]

The state $(K_0^{i+1},\ldots,K_7^{i+1})$ is derived from $(K_0^i, \ldots, K_7^i)$ by performing
the following operations:

\begin{itemize}
    \item First, apply the linear maps $M_0$ and $M_1$ to two pairs of registers. These are invertible
    linear transformations over 64 bits, acting on two 32-bit words $(a, b)$, and are defined as follows:
    \begin{align*}
        M_0(a, b) &= \big(S_{32}^1(a) \oplus b,\ S_{32}^3(b) \oplus S_{32}^1(a) \oplus b\big), \nonumber \\
        M_1(a, b) &= \big(S_{32}^9(a) \oplus b,\ S_{32}^{28}(b) \oplus S_{32}^9(a) \oplus b\big),
    \end{align*}
    where $S_{32}^m$ denotes the left circular shift of a 32-bit word by $m$ positions.

    \item If $i$ is even, apply $M_0$ to the register pairs $(K_0^i, K_1^i)$ and $(K_4^i, K_5^i)$.

    \item If $i$ is odd, apply $M_1$ to the register pairs $(K_2^i, K_3^i)$ and $(K_6^i, K_7^i)$.

    \item Next, apply a word-level permutation to the updated state:
    \[
    \begin{aligned}
    P_0: &\ (K_1 \leftrightarrow K_2),\ (K_5 \leftrightarrow K_6), \ \text{for even } i, \\
    P_1: &\ (K_1 \leftrightarrow K_4),\ (K_3 \leftrightarrow K_6), \ \text{for odd } i.
    \end{aligned}
    \]

    \item Finally, a round-dependent counter is XORed into register $K_7$ to eliminate structural
    symmetries across rounds.
\end{itemize}


\section{Symbolic Modeling of the \Aradi\ Cipher}

In this section, we present a complete symbolic model of the \Aradi\ block cipher over the Boolean field $\FF = \GF(2)$. 
The 128-bit plaintext and the 256-bit master key are represented as symbolic variables, while each round key bit
is expressed as a linear polynomial in the master key variables, in accordance with the key schedule algorithm.

At the beginning of the first round, the 128-bit internal state, consisting of four 32-bit words, is initialized
directly from the plaintext variables and updated by XORing with the 128-bit round key. This results in an internal
state described by 128 linear polynomials in the plaintext and master key variables.

The updated state proceeds through the nonlinear S-box layer, where we introduce 128 new symbolic variables
to represent the outputs of the 32 parallel 4-bit S-boxes. Each S-box is constrained by 21 quadratic polynomials
relating its four input polynomials to its four output variables.

Following the S-box layer, the linear diffusion layer is represented by 128 linear polynomials, each expressing
an output bit as a linear combination of the 128 S-box output variables.

In subsequent rounds, the internal state is updated by XORing the round key with the output of the previous round's
diffusion layer. Each round repeats the modeling of the S-box and diffusion layers in the same way: the S-box layer
introduces new symbolic output variables constrained by quadratic polynomials, and the diffusion layer
represents its output bits as linear polynomials in the S-box output variables.

This cycle of key addition, nonlinear substitution, and linear diffusion repeats for all 16 rounds. At the end
of the 16th round, the final 128 linear polynomials directly represent the ciphertext bits. Together with
the quadratic polynomials generated in each round, they form a complete symbolic representation of the cipher’s
algebraic structure. Substituting the known plaintext values and equating the final linear polynomials to the known
ciphertext bits yields a fully instantiated system of quadratic and linear equations in the master key
variables and the intermediate variables introduced at each S-box layer.

The multistep solving strategy described in the previous sections is applied to the above polynomial system
and the corresponding Boolean field equations, as a tool for algebraic cryptanalysis. The following notation
and setup provide the basis for the formal round-by-round description of the symbolic model.

\subsubsection*{Notation and Setup}

We work over the Boolean field $\FF = \GF(2)$ and consider the multivariate polynomial algebra
\[
R = \FF[k_0, \ldots, k_{255}, W^0, X^0, Y^0, Z^0, S^i_W, S^i_X, S^i_Y, S^i_Z\ (1\leq i\leq 16)]
\]
where:
\begin{itemize}
  \item $k_0, \ldots, k_{255}$ are the 256 master key variables;
  \item $W^0, X^0, Y^0, Z^0$ denote the four 32-bit words of the plaintext;
  \item $S^i_W, S^i_X, S^i_Y, S^i_Z$ ($1\leq i\leq 16$) denote the 4 S-box layer output words
  of 32 symbolic variables each, introduced in round $i$.
\end{itemize}

For each round $i$, the 128-bit round key is denoted by $RK^i = (K_W^i, K_X^i, K_Y^i, K_Z^i)$,
where each $K_T^i$ ($T = W,X,Y,Z$) is a vector of 32 linear polynomials in the 256 master
key variables:
\[
\begin{array}{l@{\,}c@{\,}ll@{\,}c@{\,}l}
K_W^i & = &(k_{w,0}^i, \dots, k_{w,31}^i), & K_X^i & = & (k_{x,0}^i, \dots, k_{x,31}^i), \\
K_Y^i & = & (k_{y,0}^i, \dots, k_{y,31}^i), & K_Z^i & = & (k_{z,0}^i, \dots, k_{z,31}^i).
\end{array}
\]
To describe the round-by-round evolution of the internal state, we denote by $W^i, X^i$, $Y^i, Z^i$
the symbolic 32-bit words at the input of round $i$, represented as polynomials in $R$.

We now describe the symbolic modeling of each round's operations within the polynomial algebra $R$.

\subsection*{Round Key Addition}

At the beginning of round $i$, one has that the internal state $(W^i, X^i, Y^i, Z^i)$, each a 32-bit
word represented as polynomials in $R$, is updated by XORing with the corresponding round key
$(K_W^i, K_X^i, K_Y^i, K_Z^i)$. Explicitly, for each bit index $0\leq j\leq 31$, it holds
\[
\bar{w}^i_j = w^i_j + k^i_{w,j},\, \bar{x}^i_j = x^i_j + k^i_{x,j},\,
\bar{y}^i_j = y^i_j + k^i_{y,j},\, \bar{z}^i_j = z^i_j + k^i_{z,j}.
\]
The resulting polynomials $\bar{w}^i_j, \bar{x}^i_j, \bar{y}^i_j, \bar{z}^i_j\in R$ form the input
to the nonlinear substitution layer.

\subsection*{Nonlinear Substitution}

For each index $0\leq j\leq 31$ in round $i$, the 4-bit input to the S-box layer is
\[
(x_0, x_1, x_2, x_3) = (\bar{w}^i_j, \bar{x}^i_j, \bar{y}^i_j, \bar{z}^i_j),
\]
and the 4-bit output is
\[
(y_0, y_1, y_2, y_3) = (s^i_{w,j}, s^i_{x,j}, s^i_{y,j}, s^i_{z,j}).
\]
The nonlinear substitution is modeled, for each S-box, by 21 quadratic polynomials in the polynomial
algebra $R$, which implicitly define the S-box transformation $\textsf{Sbox}(x_0, x_1, x_2, x_3) =
(y_0, y_1, y_2, y_3)$, namely
\begin{multline*}
x_0 x_2 + x_1 x_2 + x_3 + y_3,\ x_0 x_1 + x_0 x_2 + x_0 y_1,\ x_0 x_2 + x_0 y_2 + x_0 y_3,
\ x_0 x_2 + x_0 x_3 \\
+ x_1 y_1 + x_0 y_2 + y_1,\ x_0 x_3 + x_1 x_3 + x_0 y_0 + x_0 y_2 + x_1 y_2 + y_0,\ 
x_0 x_2 + x_0 x_3 + x_1 x_3 \\
+ x_0 y_2 + x_1 y_3 + x_3 + y_3,\ x_0 x_2 + x_1 + y_1,\ x_0 x_2 + x_0 y_2 + x_2 + y_2,\ 
x_2 y_1 + x_3 + y_3, \\
x_0 x_2 + x_0 x_3 + x_1 x_3 + x_0 y_2 + x_2 y_2 + x_0 + x_2 + x_3 + y_0 + y_3,\
x_2 x_3 + x_2 y_3 + x_3 + y_3, \\
x_0 x_2 + x_0 x_3 + x_2 y_0 + x_3 y_0 + x_0 + y_0,\ x_3 y_1 + x_0 + x_3 + y_0 + y_3,\
x_0 x_1 + x_0 x_2 + x_2 x_3 \\
+ x_1 y_0 + x_0 y_2 + x_3 y_2 + x_0 + y_0,\ x_0 x_2 + x_2 y_0 + x_3 y_3 + y_3,\
x_0 x_1 + x_0 x_2 + y_0 y_1 \\ 
+ x_0 + y_0,\ x_0 x_2 + x_0 y_0 + x_2 y_0 + x_0 y_2 + y_0 y_2 + x_0,\
x_0 x_2 + x_0 y_2 + y_0 y_3 + x_0 + y_0, \\ 
x_0 y_0 + y_1 y_2 + x_0 + x_3 + y_3,\ y_1 y_3 + x_0 + y_0,\ x_0 x_2 + x_2 x_3 + x_0 y_2 +
y_2 y_3 + x_3 + y_3.
\end{multline*}
Altogether, the 32 parallel S-boxes contribute a total of 672 quadratic polynomials in $R$ for round $i$.

\subsection*{Word-wise Linear Transformation}

The linear diffusion layer applies the transformation $\mathsf{L}_{(a_i,b_i,c_i)}$ to each 32-bit
word produced by the S-box layer. Consider, for example, one such word, say
$S^i_W = (s^i_{w,0}, \ldots, s^i_{w,31})$, and split it into two 16-bit symbolic halves:
\[
l^i = (s^i_{w,0}, \ldots, s^i_{w,15}),\ r^i = (s^i_{w,16}, \ldots, s^i_{w,31}).
\]
The transformation $\mathsf{L}_{(a_i,b_i,c_i)}(S^i_W)$ produces new linear polynomials in $R$ with
the updated halves $(l^i_0, \ldots, l^i_{15})$ and $(r^i_0, \ldots, r^i_{15})$ defined for each
$0\leq j\leq 15$ by:
\[
\begin{aligned}
l^i_j &= l^i_j + l^i_{(j + a_i) \bmod 16} + r^i_{(j + c_i) \bmod 16}, \\
r_j &= r^i_j + r^i_{(j + a_i) \bmod 16} + l^i_{(j + b_i) \bmod 16}.
\end{aligned}
\]
where the addition in subscripts is ordinary integer addition, while all other additions are in the
polynomial algebra $R$ over the base field $\FF = \GF(2)$. The constants $(a_i,b_i,c_i)$ are
specified in Table~\ref{table:shifts}.

This transformation $\mathsf{L}_{(a_i,b_i,c_i)}$ is applied independently to the four symbolic
S-box output words, yielding the updated four 32-bit symbolic words of the internal state
for the next round: 
\[
\begin{aligned}
W^{i+1} &= \mathsf{L}_{(a_i,b_i,c_i)}(S^i_W), &\ 
X^{i+1} &= \mathsf{L}_{(a_i,b_i,c_i)}(S^i_X), \\
Y^{i+1} &= \mathsf{L}_{(a_i,b_i,c_i)}(S^i_Y), &\
Z^{i+1} &= \mathsf{L}_{(a_i,b_i,c_i)}(S^i_Z).
\end{aligned}
\]
These updated words consist of linear polynomials in the algebra $R$, and no new symbolic
variables are introduced at this step. 

Each round of the \Aradi\ cipher is symbolically modeled by quadratic polynomials from the S-box
layer and linear polynomials from the key addition and diffusion layers, all defined over
the multivariate polynomial algebra $R$. Since the output of the diffusion layer becomes the input to the
next round’s key addition, and the resulting expressions feed into the nonlinear S-boxes,
the full symbolic model of the 16-round cipher comprises $16 \times 672$ quadratic polynomials
from the S-box layers, along with 128 linear polynomials from the final diffusion layer.
Altogether, the system consists of 10,880 multivariate polynomials in $R$, involving 2,432 variables:
256 master key bits, 128 plaintext bits, and $128\times 16$ intermediate S-box output bits.
These polynomials collectively define an implicit algebraic representation of the \Aradi\ cipher.

For any known plaintext-ciphertext pair, substituting the known values results in a fully
instantiated system of polynomial equations in the master key variables and the intermediate S-box
output variables. Along with the Boolean field equation ideal $\cF$ corresponding to the polynomial
algebra $R$, this system forms the input for the multistep polynomial system solver
used in a key-recovery attack, which we discuss in the following sections.

We note that the final key addition layer of the \Aradi\ encryption is not considered in this
symbolic representation, as its linear nature does not affect the algebraic complexity relevant
to our key-recovery attack.


\section{An algebraic attack to \Aradi\ based on the multistep strategy}

This section explains how the multistep solving strategy can be applied to perform an algebraic attack on the
full-round \Aradi\ cipher. Given a known plaintext-ciphertext pair, the plaintext variables in the system
are assigned fixed values. The 128 linear polynomials produced by the final round's linear layer are then
equated to the corresponding ciphertext bits, resulting in a system of multivariate polynomial equations.

The system involves $256$ master key variables and $2048 = 16\times 128$ auxiliary variables, representing
the outputs of the 32 parallel 4-bit S-boxes across $16$ rounds. In total, it comprises $16 \times 672 = 10,752$
quadratic equations arising from the S-box layers, along with $128$ final linear equations, yielding
a total of $10,880$ equations in the multivariate polynomial algebra
\[
R = \FF[k_0, \ldots, k_{255}, W^0, X^0, Y^0, Z^0, S_W^i, S_X^i, S_Y^i, S_Z^i\ (1\leq i\leq 16)].
\]
where $\FF = \GF(2)$. All computations for solving the system are performed modulo the Boolean field equation
ideal $\cF\subset R$.

Our goal is to recover the values of the unknown key bits $k_0, \ldots, k_{255}$ by solving this system.
Direct \Gr\ basis computation on the full system is computationally infeasible. In particular, matrix-based algorithms
such as F4, F5 or XL \cite{CKPS,F4,F5} suffer from exponential growth in the number of monomials and matrix dimensions.
This quickly leads to memory exhaustion, even when using sparse matrix techniques.

To address this, the classical hybrid solving strategy is often employed in algebraic cryptanalysis. However,
the number of variables that must be exhaustively guessed is typically large in order to make all resulting
polynomial systems feasibly solvable. Moreover, this number is generally determined empirically, requiring
numerous attempts with different number of variables and guesses to identify a workable configuration.

The multistep strategy offers a more automated algorithm that, starting from a reasonable initial set of guessed
variables, explores the algebraic structure of the polynomial system and extends this set only when strictly necessary.

The multistep solving strategy, namely the \MultiSolve\ algorithm, has been implemented in the computer algebra
system \Magma\ \cite{MAGMA}, utilizing its built-in \Gr\ basis computation engine. Given the high sparsity
of the polynomial systems involved in the algebraic attack on \Aradi, we make use of the sparse variant
of the F4 algorithm.

The multivariate polynomial system representing the full 16-round attack on \Aradi\ is derived according to
the symbolic modeling described in Section 5.

To evaluate the effectiveness of the multistep strategy, we performed a key-recovery attack using a random
plaintext-ciphertext pair derived from a randomly generated master key. Each experiment begins by generating
a random master key, selecting a random plaintext, and computing the corresponding ciphertext. Subsequently,
a random initial assignment is generated for a subset of the master key variables. Using this information,
the associated polynomial system is instantiated, and the \MultiSolve\ algorithm is applied to attempt
its resolution. Since the initial guesses for some master key bits are random, and therefore likely incorrect,
all the polynomial systems generated along the multistep strategy are inherently inconsistent.

To construct the initial subset of variables to be guessed, the 256 key variables are first partitioned
into 8 blocks of 32 variables each, namely
\[
B_l = \{k_{32 l+i}\mid 1\leq i\leq 32\}\ (0\leq l\leq 7).
\]
Various permutations of these blocks are then generated. For each permutation, a subset of $k$ variables
is selected by taking the first $k$ variables from the concatenated sequence of the permuted blocks.
By empirical analysis, the fastest choice appears to be the reverse permutation
\[
B_7\cup B_6\cup \ldots B_1\cup B_0.
\]

We developed a comprehensive and versatile implementation of the \MultiSolve\ algorithm within the \Magma\
computer algebra system. The implementation supports a range of oracle functions, allowing for flexible
experimentation with different heuristics and decision strategies throughout the solving process.

Through a series of experiments on \Aradi, we identified effective parameter settings for \MultiSolve\
in the context of this attack: specifically, $d_1 = 0$, which disables \GBElimLin, and $d_2 = 2$ for \GBSafe.
This choice is motivated by the fact that the input polynomials are quadratic, which enables \GBSafe\
to be applied efficiently at this degree while still producing complete \Gr\ bases, thus rendering \GBElimLin\
unnecessary.

For wild cases, we set a time limit of $\tau = 60$ seconds. The oracle function employed in \MultiSolve\ is
\OracleT. After conducting several tests with an initial number of variables set to 240, it became clear that
tame cases always occur when the number of guessed key variables is 251 or 252. To obtain the classic hybrid
strategy, it was therefore necessary to begin with 252 evaluations. Our experiments also confirmed that
assigning correct values to 251 or 252 key variables is sufficient to recover the master key.

A total of $2^{16}$ experiments were performed, each using a distinct random master key, plaintext, and partial
assignment of key bits. For \MultiSolve, 99.98\% of the cases were tame with 251 assignments. The cost of identifying
the remaining 0.02\% of wild cases via timeout and resolving them with 252 key variable evaluations was therefore
negligible.

We report here the average runtimes of \GBSafe\ in solving tame cases, along with the corresponding standard
deviations to highlight the stability of the results. All timings are given in seconds. All computations were
performed on a Linux-based system running Ubuntu 22.04 LTS, equipped with an AMD EPYC 7763 64-core processor
and x86\_64 architecture. For \Gr\ basis computations, we used the sparse F4 algorithm as implemented in \Magma\
version 2.27.5.

\begin{table}[ht]
\centering
\caption{GB Average Runtime for Tame Cases}
\begin{tabular}{|c|c|c|}
\hline
\textsf{\# Guessed Vars} & \textsf{Solving Time (s)} & \textsf{Std.~Dev.~(s)} \\
\hline
$251$ & 34.629 & 1.892  \\
$252$ & 24.957 & 1.756  \\
\hline
\end{tabular}
\end{table}

Since $\log_2(34.62) = 5.11$ and $\log_2(24.95) = 4.64$, the overall complexity of the multistep strategy is
approximately $2^{256}$ seconds, while that of the classical hybrid strategy is around $2^{256.5}$ seconds.
Given that a single encryption in a brute force attack takes significantly less than one second, we cannot claim
that \Aradi\ is broken. However, our attack has revealed some structural features of the symbolic model.

Indeed, our experiments show that assigning 251 key variables leads to a substantial reduction in the complexity
of solving the derived quadratic polynomial systems. Only a small fraction of the instances require 252 assignments,
and all systems are solved in approximately 30 seconds with a solving degree of two, equal to the input degree.
This behavior is likely due to the extreme sparsity of the corresponding Macaulay matrices and the fact that
these polynomial systems deviate significantly from the semi-regular case. Notably, the systems involve around
2000 variables.

We also experimented with various combinations of the parameters $d_1,d_2$ and $\tau$, but none of these
configurations enabled us to obtain tame cases with only 250 key variable evaluations.
In other words, fixing 251 key variables in \Aradi’s symbolic model results in a marked drop
in computational effort.


\section{Conclusions and further directions}

In this paper, we propose a new version of the multistep solving strategy, based on a recursive Depth-First
Search implementation and on the introduction of an ``oracle function'' intended to predict whether
a polynomial system requires further simplification through variable evaluations.

This approach enables a simple yet effective implementation of the \MultiSolve\ algorithm and supports
a complexity analysis based on tree structures. Furthermore, it provides a unified framework that
generalizes various solving strategies, including the one used in our attack on the \Trivium\
cipher and the classic hybrid approach, into a single algorithm, where only the oracle function needs
to be adapted. Thanks to this work, we were able to perform an algebraic attack on the recent block cipher \Aradi,
which revealed some aspects of its symbolic modeling.

As a further direction for this line of research, we suggest developing oracle functions based on machine
learning algorithms, trained on the ideals and corresponding Macaulay matrices generated by \MultiSolve.
A preliminary experimental study is currently underway. Another promising direction lies in the development
of a parallel implementation of the \MultiSolve\ algorithm, as well as in its extension to polynomial systems
admitting multiple solutions, which could be leveraged in discrete optimization tasks.

\section*{Acknowledgments}
The authors would like to thank the anonymous reviewers for their careful reading of the manuscript
and for their constructive comments, which enhanced the clarity and readability of the work.

\end{document}